\documentclass[conference]{IEEEtran}
\IEEEoverridecommandlockouts
\usepackage{cite}
\usepackage{amsmath,amssymb,amsfonts}
\usepackage{algorithm}
\usepackage{algpseudocode}
\usepackage{graphicx}
\usepackage{textcomp}
\usepackage{xcolor}
\usepackage{xspace}
\usepackage{array}
\usepackage{url}
\usepackage{balance}

\def\BibTeX{{\rm B\kern-.05em{\sc i\kern-.025em b}\kern-.08em
    T\kern-.1667em\lower.7ex\hbox{E}\kern-.125emX}}
\begin{document}

\newcommand{\methodname}{ScenarioForge\xspace}

\title{\methodname: Resource-Grounded Generation and Execution Validation of Cyber Exercises}

\author{\IEEEauthorblockN{Jaime C. Acosta}
\IEEEauthorblockA{\textit{DEVCOM Army Research Laboratory} \\
jaime.c.acosta.civ@army.mil}

\and
\IEEEauthorblockN{Anantaa Kotal}
\IEEEauthorblockA{\textit{The University of Texas at El Paso} \\
akotal@utep.edu}
\and
\IEEEauthorblockN{Aritran Piplai}
\IEEEauthorblockA{\textit{The University of Texas at El Paso} \\
apiplai@utep.edu}
}
\maketitle

\begin{abstract}
Cyber-range and capture-the-flag scenario construction remains labor-intensive because authors must coordinate infrastructure, vulnerable services, attack paths, artifacts, and success conditions. LLMs can propose detailed exercises, but plausible descriptions do not guarantee deployable software or verifiable objectives. We present \methodname, which turns high-level requirements into resource-grounded topologies, services, traffic, segmentation, and challenge chains, then instantiates and validates them in the Common Open Research Emulator (CORE). The ScenarioForge Execution Dataset contains 226 materialized scenarios covering 294 Vulhub images and all 85 custom-built challenge-resource generators developed with our AI-assisted scaffolding. All 226 passed construction, preview, sequencing, deployment, strict runtime validation, and live artifact checks; their 867 steps---each an individually verifiable challenge objective---comprise 457 vulnerability-backed and 410 generated-resource-backed instances. In the separate 50-prompt Matched AI-Authoring Benchmark, ScenarioForge+Qwen3.8 and ScenarioForge+GPT5.6 each achieved 50/50 (100\%) topology/build success and 160/160 intent fields; direct frontier generation reached 41/50 builds and 147/160 fields only with the full cumulative scaffold: structure, vocabulary, policy, and container-image guidance. The test still omitted ScenarioForge's downstream challenge-resource generation, dependency/flag sequencing, deployment, and live validation. The artifacts support operationally rich, reproducible exercises for training and evaluation.
\end{abstract}


\section{Introduction}

Cyber ranges reproduce networked systems, vulnerabilities, and adversarial activity for controlled training and evaluation. Effective exercises require more than a narrative: authors must coordinate infrastructure, vulnerable components, network dependencies, participant actions, and success conditions. This combination of domain expertise and detailed configuration makes scenario construction expensive; automation can improve diversity and adaptability \cite{nakata2020vulnerability,costa2020vsdl}.

LLMs can generate attack narratives and exercise designs, and recent work uses multiple model roles and retrieval augmentation to improve their detail and realism \cite{yamin2024applications}. Separately, cyber-range research provides executable specifications such as VSDL \cite{costa2020vsdl} and emulation platforms such as CORE \cite{ahrenholz2008core}.

The unresolved problem is the boundary between description and execution. An LLM may propose unavailable implementations, incompatible versions, invalid configurations, or unsupported objectives. Assessments of soundness, realism, creativity, or usability do not establish that a scenario can be instantiated and operated \cite{yamin2024applications,palumickas2026ascert,hasegawa2026llm}. The model's expressed \emph{knowledge} must therefore be reconciled with the range's \emph{executable resources}.

\begin{figure*}[t]
    \centering
    \includegraphics[width=0.92\textwidth]{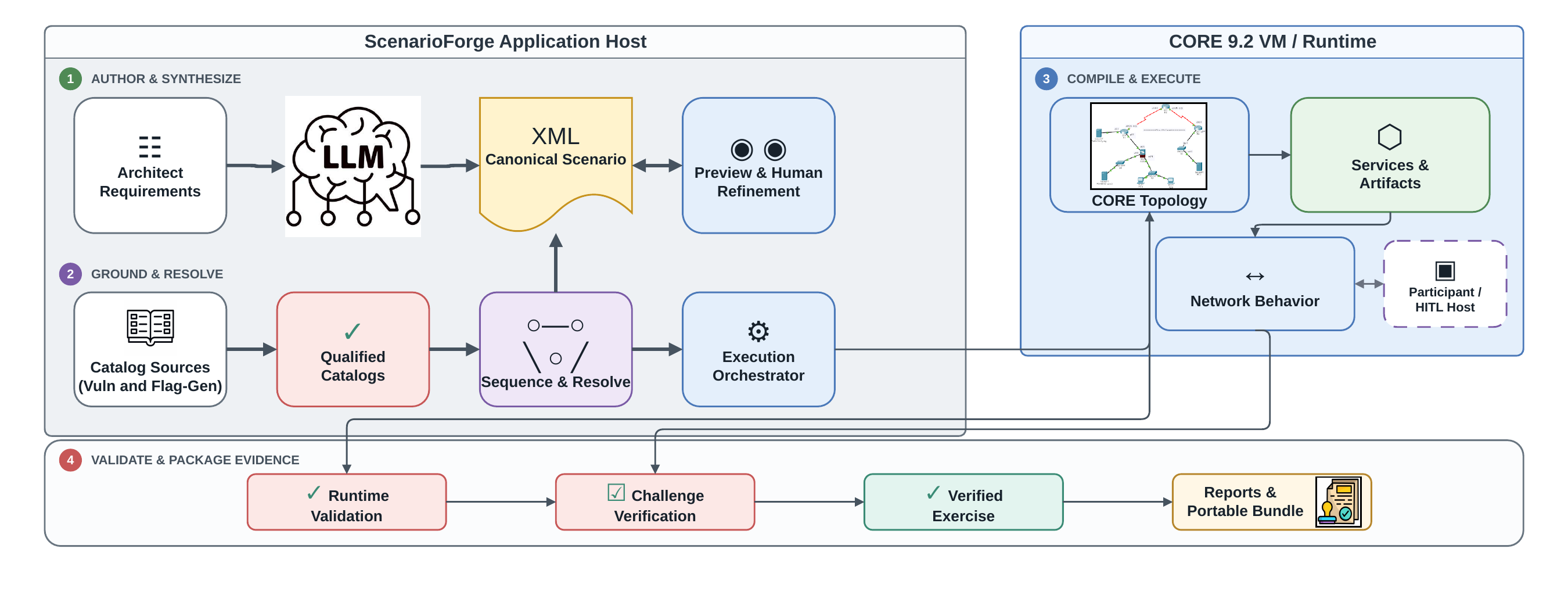}
    \caption{ScenarioForge workflow and trust boundaries. The canonical XML is refined and grounded against qualified catalogs before orchestration through CORE gRPC and SSH/SFTP paths. ScenarioForge then validates observed runtime state and challenge outcomes, producing a verified exercise and portable evidence bundle; participant or test hosts attach through HITL when configured.}
    \label{fig:system-overview}
\end{figure*}

We present \methodname (Fig.~\ref{fig:system-overview}), a framework that treats LLM output as a proposal rather than a finished environment. A typed, persistent representation separates generative authorship from deterministic grounding and execution: the model proposes scenario intent and relationships, while backend components decide which installed assets can realize them and record the resulting bindings. The resolved specification is then instantiated in CORE and judged using observations from the deployed environment. This division establishes an explicit trust boundary---plausible generated content is never accepted as evidence of an available resource, an operational service, or a supported objective.

Our contributions are:
\begin{itemize}
    \item \textbf{Automated, resource-grounded construction:} \methodname grounds topology, services, segmentation, traffic, artifacts, and challenge chains in deployable resources. It supports Vulhub environments and user-developed vulnerabilities, services, and artifacts, with built-in automated resource construction using local or cloud LLMs.

    \item \textbf{Execution-based validation and infrastructure:} Catalog preflight, preview, sequencing, deployment, runtime validation, and live artifact checks produce operational evidence. Our evaluation harness and hardened CORE 9.2 VM/fork make these checks repeatable for container-backed ranges.

    \item \textbf{A matched AI-authoring benchmark:} Across 50 separate prompts in the Matched AI-Authoring Benchmark, both ScenarioForge+Qwen3.8 and ScenarioForge+GPT5.6 achieved 50/50 (100\%) topology/build success and 160/160 requested intent fields. Thus, the relatively small, open-weight Qwen3.8 model matched the frontier model inside \methodname, and the framework-assisted conditions outperformed direct frontier generation, which peaked at 41/50 builds and 147/160 fields despite cumulative scaffolding. We release the frozen prompts and expectations, retained outputs, scoring records, and analysis scripts needed to reproduce the comparison.

    \item \textbf{A reproducible execution dataset:} We release the ScenarioForge Execution Dataset: 226 fixed scenarios covering 294 Vulhub images, all 85 custom-built challenge-resource generators, and 867 challenge steps across varied topologies, services, policies, traffic, artifacts, and dependencies.
\end{itemize}

Together, the Matched AI-Authoring Benchmark isolates description-to-buildable-definition conversion, while the Execution Dataset follows fixed, grounded definitions through sequencing, deployment, and live evidence. This proposal-to-evidence design makes the contribution more than a prompt or generator. It localizes failures to authoring, resolution, compilation, deployment, runtime state, or artifact verification; recorded phase timings characterize reproduction cost in the stated environment.

\section{Related Work}

Cyber-range automation spans reusable infrastructure, randomized generation, model-driven exercise execution, and declarative scenario languages \cite{beuran2018cytrone,yamin2020cyberranges,nakata2021cyexec,yamin2022modeling,costa2020vsdl,martou2022generating}. CRACK similarly verifies manually authored TOSCA blueprints and delegates VM and network realization to an external ARIA/OpenStack stack, but its released model does not cover containerized vulnerability catalogs or CTF-specific artifact and pivot dependencies \cite{russo2020crack}. SecGen provisions randomized vulnerable VMs and CTF challenges with generated hints from manually authored constraints \cite{schreuders2017secgen}. Automated attack execution likewise provides strong evidence when objectives are defined over known mechanisms \cite{faeroy2023automatic}.

AI-assisted work generates exercise content through information extraction, multi-agent prompting, and retrieval augmentation \cite{zacharis2023aicef,yamin2024applications}. ASCERT evaluates interactive LLM-generated scenarios for incident fidelity and repeatability, while Hasegawa and Takakura generate organization-specific exercises from real attack data \cite{palumickas2026ascert,hasegawa2026llm}. These systems broaden semantic authorship, but their reported evaluations center on generated content rather than deployment of the corresponding resources.

\methodname assigns distinct responsibilities across generation, grounding, and execution: the LLM proposes topology and exercise logic; catalogs bind proposals to implementations; CORE and Docker materialize the range; and backend observations judge supported objectives. Unlike CRACK's externally realized VM model, this path runs natively or with a standalone CORE VM and does not require OpenStack. Containerized services coexist with routed emulation, while flags, hints, solutions, typed artifact prerequisites, and access constraints remain in the executable model.

\section{Methodology}
\subsection{System Overview}

As Fig.~\ref{fig:system-overview} shows, \methodname combines LLM-based synthesis with deterministic grounding, execution, and validation. Inputs may range from a broad objective to explicit topology, service, vulnerability, routing, segmentation, access, and challenge constraints.

The canonical scenario XML persists throughout the workflow. Grounding maps network elements to CORE configurations, vulnerable services to Docker-backed resources, and CTF objectives to installed challenge-resource generators. The resolved scenario runs natively or in a CORE 9.2 VM controlled through gRPC and SSH; a participant may attach through hardware-in-the-loop networking. Runtime checks cover nodes, containers, services, ports, injects, segmentation, traffic, and artifacts. A Solutions Script traverses the challenge sequence and records pass, fail, inconclusive, or skip. XML, logs, summaries, and reports preserve the evidence.

\subsection{Scenario Specification and Intermediate Representation}

\methodname uses one XML document across authoring, preview, sequencing, and execution; when invoked, the LLM refines it but never invokes CORE, Docker, or remote shells. The representation records infrastructure, node and resource configuration, challenge state and dependencies, participant connectivity, and execution metadata. Partially specified scenarios preserve user constraints while the model fills missing fields, and every generated value must resolve to a catalog entry or backend capability. Embedded \texttt{PlanPreview} and sequencing state make the previewed topology and challenge structure the deployment inputs rather than a separate plan.

\subsection{LLM-Guided Scenario Synthesis}

The LLM extracts constraints, decomposes the request into infrastructure and exercise components, completes unspecified roles and relationships, and serializes the result into constrained XML. Explicit constraints remain authoritative, so a fixed vulnerability, node count, or challenge order is preserved while surrounding topology and services may be generated.

The WebUI combines model-guided authoring with a topology-intent compiler that preserves explicit requests. Its built-in AI-assisted Generator Builder uses local or cloud LLMs to scaffold or refine user-developed vulnerability and challenge resources as \texttt{flag-generator} or \texttt{flag-node-generator} implementations, test them against their contracts, and repair failures before installation. This workflow produced the 85 custom-built challenge-resource generators evaluated here; installed generators execute deterministically.

Because services, versions, generators, or attack steps may be unavailable or incompatible, initial output remains a proposal. Deterministic resolution establishes which concrete assets and parameters can realize it, preserving flexible authoring without delegating deployment decisions to the model.

This hybrid path is important when requests mix exact and open-ended requirements. The intent compiler converts explicit counts and named services into typed constraints, while model calls fill relational structure and unspecified values. Resolution can add the hosts required to place selected containers, reconcile identifiers after topology edits, and reject catalog choices whose architecture or runtime prerequisites do not match the execution endpoint. Validation and repair operate on the persisted representation, so a corrected draft is rechecked against the same constraints rather than regenerated from an unconstrained conversation.

\subsection{Grounding in Executable Cybersecurity Resources}

Grounding retains an element only when it maps to an executable asset or supported deterministic configuration. Vulnerable services resolve to cataloged Docker Compose environments, ports, startup procedures, and inject plans, including user-developed resources and reproducible Vulhub environments \cite{vulhub2026}. Challenge-resource generators produce artifacts, flags, deployment instructions, and solutions; network and participant resources compile to CORE, HITL, or Proxmox-supported configurations. Thus a service name alone is insufficient without a compatible version and runtime definition, and an objective is accepted only when its installed generator supplies both an artifact and a verification procedure. Static preflight rejects invalid catalog or inject wiring before execution.

The active challenge-resource generator catalog covers binary artifacts such as firmware and packet captures; credentials and signatures; JSON, YAML, and INI configurations; encoded payloads; and textual evidence such as chat exports, tickets, operations notes, and backups. Each entry records its producer, placement, expected output, and solution procedure. This makes catalog state, rather than a model's recollection, the authority for what the range can deploy and verify.

Table~\ref{tab:grounding-evidence} generalizes these bindings across component classes and distinguishes preflight acceptance from observed execution evidence.

\begin{table*}[!t]
\caption{Grounding and execution evidence across \methodname component classes}
\label{tab:grounding-evidence}
\centering
\footnotesize
\renewcommand{\arraystretch}{1.12}
\begin{tabular}{p{0.16\textwidth} p{0.34\textwidth} p{0.40\textwidth}}
\hline
\textbf{Component} & \textbf{Grounding and preflight} & \textbf{Execution evidence} \\
\hline
Vulnerable service & Resolve to a validated catalog entry with a deployment definition, exposed ports, startup procedure, and inject plan. & Container is placed on the intended node; required processes and ports are active; exploit prerequisites and injected files are present. \\
Generated challenge & Bind to an installed flag-generator or flag-node-generator that produces an artifact, expected flag, and solution procedure. & Output and inject artifacts exist at the designated target; the prescribed solution action retrieves the expected flag. \\
Topology and segmentation & Compile nodes, links, addresses, routes, and policy rows into CORE-supported operations; reject invalid references or dependency cycles. & Expected nodes and paths exist; allowed communication succeeds and prohibited communication is denied. \\
Traffic workload & Resolve protocol, endpoints, payload type, rate, and timing to supported traffic-agent configurations. & Sender and receiver agents are active and the requested flows traverse the instantiated topology. \\
Participant attachment & Bind the participant requirement to a configured hardware-in-the-loop interface and attachment target. & The participant host reaches permitted entry points through the selected CORE bridge or interface. \\
Challenge dependency & Resolve each prerequisite and produced artifact to a concrete prior or current step, including any required credential, host access, or pivot. & The Solutions Script traverses the dependencies in order, automatically checks supported steps, and records pass, fail, inconclusive, or skip. \\
\hline
\end{tabular}
\end{table*}

Algorithm~\ref{alg:resource-grounded-execution} formalizes the central deterministic pipeline. It treats the LLM output as a proposal, returns typed failures when a component or dependency cannot be realized, and produces execution evidence only after deployment and objective verification succeed. A user edit to the persisted preview re-enters the pipeline at the resolution step so that the revised XML is checked under the same rules.

\begin{algorithm}[!t]
\caption{Resource-Grounded Execution and Verification}
\label{alg:resource-grounded-execution}
\begin{algorithmic}[1]
\Require Scenario XML $X$, catalogs $C$, execution backend $B$
\Ensure Evidence record $E$ or typed failure $F$
\State $Q \gets \Call{ParseAndPreserveConstraints}{X}$
\ForAll{component $q \in Q$}
    \State $A_q \gets \Call{CompatibleAssets}{q,C,B}$
    \If{$A_q = \emptyset$}
        \State \Return $\Call{GroundingFailure}{q}$
    \EndIf
    \State $a_q \gets \Call{Select}{A_q}$
    \State $X \gets \Call{Bind}{X,q,a_q}$
\EndFor
\State $G \gets \Call{DependencyGraph}{X}$
\If{$G$ is cyclic or has an unresolved prerequisite}
    \State \Return $\Call{DependencyFailure}{G}$
\EndIf
\State $T \gets \Call{CompileTopology}{X,B}$
\If{$\neg\Call{Preflight}{X,T,C}$}
    \State \Return $\Call{PreflightFailure}{X,T}$
\EndIf
\State $X \gets \Call{PersistPreview}{X,T,G}$
\State $O \gets \Call{TopologicalOrder}{G}$
\State $D \gets \Call{Deploy}{T,O,B}$
\If{$\neg D.\text{ok}$}
    \State \Return $\Call{DeploymentFailure}{D}$
\EndIf
\State $V \gets \Call{RuntimeValidate}{D,X}$
\State $S \gets \Call{VerifyObjectives}{D,G}$
\If{$\neg V.\text{ok} \lor \neg S.\text{ok}$}
    \State \Return $\Call{VerificationFailure}{V,S}$
\EndIf
\State \Return $E \gets (X,D,V,S)$
\end{algorithmic}
\end{algorithm}

For $n$ components, $c$ catalog entries, $e$ dependency edges, $t$ compiled records, and $p$ probes, Algorithm~\ref{alg:resource-grounded-execution} uses $O(|X|+nc+e+t+p)$ worst-case time and $O(n+e+t)$ space; indexing reduces time toward $O(|X|+n+e+t+p)$. The bound excludes LLM inference and treats backend requests as unit-cost operations.

Three invariants connect this algorithm to the implementation. First, every selected resource is identified by a catalog key and copied into the resolved specification, making model output insufficient as a binding. Second, all dependency endpoints must name persisted steps or typed artifacts before topological ordering. Third, deployment and validation consume the same resolved XML and stable node identifiers used by preview. These invariants prevent a plausible but unavailable component, an orphaned prerequisite, or a preview-only repair from silently entering execution. Failures therefore have typed locations: grounding, dependency analysis, preflight, deployment, runtime validation, or objective verification.

\subsection{Topology and Exercise Construction}

After grounding, \methodname compiles hosts, routers, links, addressing, routes, service placement, traffic, segmentation, and artifacts into CORE operations. Construction follows dependency order: networks precede services, targets precede injects, and artifacts follow active destinations. A preview from the execution XML exposes the exact topology, roles, vulnerabilities, and dependencies before deployment.

\subsection{Dependency-Aware Challenge Construction}

\methodname represents each exercise as a dependency graph. A \emph{step} is one verifiable challenge objective bound to a grounded service, file, credential, or challenge node and an expected result. The framework connects requirements to predecessor outputs, rejects unresolved or cyclic graphs, and records targets, flags, and solution actions. \emph{Sequence} constructs a valid order and \emph{Resolve} selects resources and parameters; both update the same XML.

Sequencing persists more than an ordinal list. Each assignment records the target node and service, the resource that implements the objective, prerequisite access facts, typed inputs produced by earlier steps, expected outputs, and the verification strategy. Topological ordering then yields a deployable Flow in which generated files, credentials, endpoints, and access capabilities become explicit edges rather than instructor-only assumptions. The persisted sequencing JSON and updated XML are consumed by execution, the Solutions Script, and artifact validation, so all three observe the same resolved dependency graph.

After deployment, the Solutions Script traverses the graph and records pass, fail, inconclusive, or skip. Supported HTTP, SSH, and TCP challenge-node strategies are automatic; vulnerability-host and artifact-only steps preserve flags, hints, solutions, and interfaces for participant or exploit-harness interaction. Intermediate outputs reach only dependent steps, while strict validation establishes deployment and artifact readiness across the resolved graph.

\subsection{Segmentation and Pivot Constraints}

Pivoting is a dependency edge from provider $H_p$ to target $H_t$: a predecessor produces access to $H_p$, and the dependent step requires it before targeting $H_t$. Providers may be vulnerable hosts, generated challenge nodes, or supported SSH fallbacks.

The graph must link each requirement to its provider; policy must permit the service from $H_p$ and deny unauthorized sources; probes require success from $H_p$ and failure from representative prohibited sources. This positive-and-negative test avoids treating allow-list membership as proof of isolation.

Each dependent step carries its provider--target pair, so multiple or non-adjacent pivots do not overwrite global context. Paths are probed independently and reported through the common verification record.

\subsection{Human-in-the-Loop Scenario Refinement}

Users may provide a broad objective or detailed requirements, or bypass AI entirely and construct the scenario in the WebUI. The shared XML exposes a resolved preview before deployment. Edits trigger re-resolution, allowing users to modify resources, placements, sequences, or connectivity without restarting authoring. When used, the model proposes intent; grounding restricts implementation, and the user retains final control.

\subsection{Cyber-Range Instantiation}

After topology, resources, and dependencies are fixed, \methodname executes natively or separates the application, CORE 9.2, and participant roles across hosts. CORE gRPC manages session topology and lifecycle; SSH handles placement, Docker, services, injects, and inspection. Execution creates the topology and policy, deploys services, applies injects and artifacts, then activates traffic and HITL attachments. WebUI and CLI workflows share this backend, and request identifiers let retried clients reconnect to long operations.

\subsection{Multi-Stage Validation and Verification}

Validation separates three failure classes. Catalog preflight checks deployment metadata, services, and inject wiring. Runtime validation inspects the active CORE session for expected nodes, containers, ports, injects, outputs, traffic, and policy behavior. Challenge verification traverses the resolved sequence, automatically exercises supported strategies, and records pass, fail, inconclusive, or skip for each step. WebUI and \texttt{check-artifacts} CLI paths return a structured \texttt{validation\_summary}; strict success requires its \texttt{ok} field and all relevant issue counters to be clear. Passing one stage does not imply success at a later one.

In particular, strict validation requires zero missing nodes, stopped Docker services, missing injects, missing generator outputs, and missing generator injects. Positive and negative reachability probes check segmentation, while artifact checks inspect the deployed environment rather than merely the generated plan. Per-issue logs retain the observation that caused a failure, preserving the boundary among catalog qualification, composed runtime state, and objective completion.

\section{ScenarioForge Execution Dataset}

\subsection{Construction and Intended Configuration}

The ScenarioForge Execution Dataset contains 226 fixed, materialized specifications, each preserving its authoring input, seed, catalog choices, authoritative XML, preview, challenge order, execution evidence, and derived composition. For every case, ScenarioForge generated and persisted Flow order, challenge dependencies, and flag placement from the resolved challenge graph; all 226 then traversed deployment, strict runtime validation, and live artifact checks. Intended configuration is preserved separately from execution logs, so success does not define intended content. The collection uses one validated catalog snapshot covering 294 enabled, self-contained Vulhub images and the complete enabled set of 85 custom-built challenge-resource generators; ten imported entries requiring external online services were excluded. This dataset is distinct from the 50-prompt Matched AI-Authoring Benchmark in Section~V.

Thirty source specifications define vulnerability-focused, artifact-focused, mixed, scale, and segmented families. The evaluation harness expands their iteration counts and seeds into 226 resolved specifications before execution, freezing bounded topology choices, resource assignments, policy and traffic settings, and validation requirements. This two-level organization preserves both the reusable family definition and the exact case that ran. It also allows a future catalog or backend revision to replay identical resolved cases or deliberately rematerialize a family and report that distinction.

\subsection{Composition and Operational Breadth}

\begin{table*}[!t]
\caption{ScenarioForge Execution Dataset configuration and coverage}
\label{tab:dataset-characterization}
\centering
\fontsize{7.25pt}{8.0pt}\selectfont
\setlength{\tabcolsep}{3.5pt}
\renewcommand{\arraystretch}{0.96}
\begin{tabular}{>{\raggedright\arraybackslash}p{0.13\textwidth} >{\raggedright\arraybackslash}p{0.18\textwidth} >{\raggedright\arraybackslash}p{0.18\textwidth} >{\raggedright\arraybackslash}p{0.42\textwidth}}
\hline
\textbf{Category} & \textbf{Measure} & \textbf{Scale / incidence} & \textbf{Composition and coverage} \\
\hline
\textbf{Dataset scope} & Scenarios & \textbf{226} specifications & Fixed and materialized; one challenge chain per scenario. \\
 & Challenge objectives (steps) & \textbf{867} total & 457 vulnerability-backed (52.7\%); 410 generated-resource-backed (47.3\%). \\
 & Chain structure & \textbf{212} multi-objective (93.8\%) & Mean 3.836 objectives per scenario. \\
 & Chain-length distribution & \textbf{1--10 objectives per chain} & Scenarios: 14 have 1 objective; 24 have 2; 36 have 3; 113 have 4; 12 have 5; 11 each have 6 or 7; 4 have 8; none have 9; 1 has 10. \\
\textbf{Resource catalog} & Vulnerability resources & \textbf{294} enabled images & 457 selections; 181/226 scenarios (80.1\%); 136 component types; 8 semantic families. \\
 & Vulnerability severity & \textbf{242/294} map to CVEs (82.3\%) & Published CVSS: 183/294 (62.2\%), mean $8.74 \pm 1.41$; 107 critical (58.5\%), 56 high (30.6\%), 20 medium (10.9\%). CISA Known Exploited Vulnerabilities (KEV): 97/242 (40.1\%). \\
 & Generated resources & \textbf{85} custom-built generators & 100\% coverage; 410 selections; 186/226 scenarios (82.3\%); 15 service types; 7 semantic families. \\
\textbf{Infrastructure} & Planned topology & \textbf{2,869} nodes & 465 routers; 1,537 Docker hosts; 867 challenge nodes. Per scenario: 5--27 nodes and 1--5 routers. \\
 & Service placement & \textbf{822} assignments & \\
\textbf{Network policy} & Segmentation & \textbf{209} policy rows & 106 firewall; 103 NAT. 148 scenarios (65.5\%) segmented. \\
 & Access providers & \textbf{54} rows & 15 generated-node; 13 vulnerability; 10 SSH fallback; 16 automatic. 42 scenarios (18.6\%) access-enabled. \\
\textbf{Traffic} & Definitions and pairs & \textbf{382} traffic rows & 532 requested source--destination pairs. \\
 & Synthetic profiles & \textbf{5} seeded profiles & Requested pairs: audio 101; gibberish 115; photo 101; text 95; video 120. \\
\hline
\end{tabular}
\end{table*}

The challenge objectives are nearly balanced between vulnerability-backed and generated-resource-backed steps (Table~\ref{tab:dataset-characterization}). Generated objectives contribute typed files, credentials, configurations, endpoints, and transcripts that support transformations and service interactions rather than independent flag retrieval. Topology, policy, access, and traffic records add routed, segmented, and source-dependent structure. Traffic labels are reproducible synthetic byte profiles, rather than complete media files; real streams can replace them without changing the scenario model.

Table~\ref{tab:dataset-characterization} also separates catalog coverage from scenario distribution: complete enabled-catalog coverage coexists with a deliberately varied, predominantly multi-objective suite. The chain distribution includes both the modal four-objective cases and observed extremes through ten objectives, so the strict result is not confined to the simplest configurations.

Beyond the aggregate family counts in Table~\ref{tab:dataset-characterization}, vulnerability resources are concentrated in web applications/frameworks/CMS, data/search/storage, and network/web infrastructure. The dataset also provides more detailed per-year vulnerability statistics, including CVE identifiers and publication-year comparisons. Generated challenges span file sharing, data stores, web delivery/application access, remote administration/source access, messaging/transfer, software supply chain, and identity/discovery. Their typed outputs include files, checksums, endpoints, tokens, source-access capabilities, and shells.

The network and policy records summarized in Table~\ref{tab:dataset-characterization} create routed, segmented, and source-dependent paths rather than isolated hosts. The synthetic traffic profiles use recognizable seeded framing---HTTP-like text, JPEG-like markers, H.264-style start codes, repeated audio frames, and random bytes---to exercise composition while keeping replay independent of external media.

\subsection{Intended Uses and Coverage}

Fixed inputs and bindings support regression testing, controlled agent or participant comparisons, and composition analysis without reconstructing intent from logs. The dataset's deliberate composition combines complete enabled-catalog coverage with varied family sizes, chain lengths, topology scales, access patterns, segmentation policies, and reproducible traffic profiles.

Each record provides three aligned views of the same exercise: source and resolved specifications capture intended configuration; preview and sequencing outputs capture compiled topology and dependency structure; and runtime logs, structured validation, and artifact reports capture observed execution. Stable scenario, node, and challenge identifiers connect these views. This organization supports separate measurement of intent preservation, topology construction, deployment, policy behavior, and objective evidence while retaining the exact configuration needed to reproduce or extend a case.

\section{Experimental Evaluation}

\subsection{Experimental Questions and Evidence Boundary}

The experiments use separate inputs. \textbf{EQ1} asks whether all 226 fixed Execution Dataset scenarios traverse XML construction, preview, sequencing, deployment, strict runtime validation, and live artifact checks. \textbf{EQ2} asks whether resource-grounded authoring produces CORE-buildable topologies and preserves requested intent better than direct frontier generation across the independent 50-prompt Matched AI-Authoring Benchmark. EQ2 measures the topology-facing items controlled by the AI Generator through safe CORE construction. \methodname's downstream sequencing stage derives Flow order, challenge dependencies, and flag sequence from the grounded topology and catalog metadata; EQ1 evaluates these stages. A separate no-Flow audit starts the framework-assisted outputs.

\subsection{Evaluation Infrastructure and Full-Pipeline Study}

We built \texttt{scenarioforge-eval} to exercise each fixed specification once. It parses YAML, persists seed and XML, invokes preview, sequencing, execution, and artifact checks, serializes phases sharing a CORE VM, and retains inputs, logs, metrics, and reproduction packages. Strict success requires normal process exit, \texttt{validation\_summary.ok}, zero issue counters, and observed expected containers, services, ports, injects, policies, traffic, and access paths. Its WebUI controls runs and indexes outcomes, timings, failure locations, topology, composition, and evidence.

The harness exposes separate Execute and Analysis views. Execute selects the specification set, ScenarioForge checkout, output directory, phase boundary, and replay mode; it streams bounded logs and supports stopping, resuming, or rerunning selected phases. Analysis indexes result trees and filters by run metadata before presenting topology, composition, timing, phase outcomes, artifacts, and validation evidence. This interface supported live diagnosis during the study, while the retained JSON and CSV records remain the source for reported measurements.

The evaluation VM packages our customized CORE 9.2 fork. We fixed container-initialization races, Docker/Podman node and interface persistence across gRPC/XML, stale interfaces, Compose project and session isolation that could cause collisions or empty volumes, and container shell/environment discovery. These are repeatability and runtime-infrastructure contributions, not AI Generator outputs.

The batch used a 16-core, 48~GB Apple M4 Max host and CORE 9.2 in a local 8~GB VMware VM. We report the 194 runs whose images were already cached separately from the 32 runs in which image pulls were observed, exposing cache state alongside each timing range.

Under this infrastructure, all 226 executions completed XML construction, preview, sequencing, deployment, strict validation, and live artifact checks with no phase failures. The specifications establish intended membership; retained batch records establish outcomes and localize future failures by phase.

Minimum--maximum times (pre-cached, $n=194$ / pull observed, $n=32$; seconds) were: XML, 0.003--0.994 / 0.298--0.830; preview, 0.443--0.703 / 0.456--0.846; sequencing, 14.9--74.3 / 15.9--58.3; execution, 56.4--657.2 / 70.8--486.8; and total, 73.7--727.0 / 87.7--550.0. These observed reproduction costs include composition, image builds, service startup, network convergence, and validation.

\subsection{Matched AI-Authoring Benchmark}

The Matched AI-Authoring Benchmark contains 50 arm-neutral requests organized into five complexity tiers, progressing from routed networks through services, segmentation, vulnerable containers, and combined stress cases. We used \texttt{gpt-5.6-sol} to generate the initial suite with this instruction: \emph{``Generate 50 concise, arm-neutral cyber-range requests in five increasing-complexity tiers containing 12, 12, 10, 10, and 6 cases. Progress from routed host networks to services, firewall/NAT segmentation, vulnerable containers, and combined stress cases. Vary router and host counts; do not mention ScenarioForge, its schema, catalog names, or tools. For each request, return machine-readable expected router, host, service, segmentation, vulnerability, and traffic fields.''} The resulting requests and expectations were frozen before any evaluated condition ran.

Scorers parse each condition's native output; missing or unreadable output remains in the denominator. Seven conditions, each evaluated against 50 prompts, yield 350 retained native authoring artifacts---100 framework-assisted ScenarioForge definitions and 250 direct response/XML records---and 1,120 intent-field decisions (160 per condition). The framework runner permits one retry only after provider timeout; reported token and timing values use the retained successful attempt.

Across the frozen suite, each condition receives requests totaling 462 topology-node instances: 128 routers (mean 2.56; range 1--5) and 334 hosts (mean 6.68; range 2--15). Every prompt specifies both counts. Twenty-four prompts request 38 named services (15 SSH, 15 HTTP, and 8 HTTPS); 18 request 24 segmentation instances (14 Firewall and 10 NAT); 15 request 29 vulnerable-service placements; and three request background traffic. Tier-specific feature totals progress from none in T1 to 19 services in T2; 3 services and 12 policies in T3; 3 services, 2 policies, and 18 vulnerabilities in T4; and 13 services, 10 policies, 11 vulnerabilities, and all three traffic requests in T5. These six field types produce the 160 intent decisions per condition. Per-request expectations and detailed tier composition are released under \path{frontier-comparison/specifications/} in the dataset repository.

The framework-assisted arm sends only the description through \methodname's AI Generator. We denote the open-weight \texttt{Qwen3.8-27B-NVFP4} condition \emph{ScenarioForge+Qwen3.8} and the frontier \texttt{gpt-5.6-sol} condition \emph{ScenarioForge+GPT5.6}; both receive the same schema, compiler, catalogs, validation, and repair path. The direct arm sends the same description to \texttt{gpt-5.6-sol} without \methodname. Its five cumulative scaffold levels, or rungs, progress from a CORE XML request (R1), to a structural example (R2), node/service vocabulary (R3), segmentation vocabulary (R4), and container-image guidance (R5). The cumulative design measures how much representation knowledge direct generation requires before it approaches the framework-assisted conditions, while the two AI Generator conditions expose the effect of changing the underlying model without changing the framework.

Every rung retains the same user request and all prior instructions, so differences across R1--R5 reflect added representation guidance rather than changing task content. R1 asks for a loadable CORE XML artifact. R2 demonstrates document structure, R3 enumerates permitted node and service terms, R4 adds the policy vocabulary needed for firewall and NAT requests, and R5 supplies container-image conventions for vulnerable services. The direct model receives no ScenarioForge catalog lookup, compiler, validation tool, or repair loop. Conversely, the framework conditions receive no rung-specific direct scaffold: they use the ordinary AI Generator path under test.

EQ2 excludes Flow/challenge-chain outcomes because downstream sequencing, not the AI Generator, derives order, dependencies, and flag sequence. Scoring them would misattribute framework decisions. Created with ScenarioForge's AI-assisted authoring/scaffolding tools, the 226 fixed EQ1 scenarios collectively exercise chains, challenge resources, artifact dependencies, access-controlled pivots, segmentation, traffic, deployment, and live validation.

\emph{Intent accuracy} is the fraction of requested fields matched after parsing each native representation. Host counts use the same topology-level definition in both arms: all materialized non-router hosts, including hosts that carry vulnerable services. \emph{Buildability} is successful CORE 9.2 construction of a nonempty session from the condition's native artifact. The scorer calls \texttt{open\_xml(start=False)}, causing CORE to parse and instantiate the topology definition while leaving nodes, containers, and physical interfaces inactive. Per-case and summary records are retained. The companion robustness audit starts the framework-assisted outputs, runs post-execution validation, and invokes live artifact checks with Flow disabled.

The scorer first extracts each condition's native XML, then derives observed topology and resource fields without normalizing the artifact through \methodname. This avoids silently repairing direct output before grading. Build and intent are reported separately: a structurally valid session can omit requested content, while a textually plausible artifact can preserve counts yet fail CORE construction. Keeping missing outputs and absent fields in their denominators prevents selective reporting of parseable cases.

\begin{table*}[!t]
\caption{Matched AI-Authoring Benchmark ($n=50$ prompts per condition)}
\label{tab:frontier-comparison}
\centering
\scriptsize
\setlength{\tabcolsep}{4pt}
\renewcommand{\arraystretch}{1.05}
\begin{tabular}{p{0.16\textwidth} p{0.29\textwidth} p{0.12\textwidth} p{0.14\textwidth} p{0.09\textwidth} p{0.07\textwidth}}
\hline
\textbf{Condition} & \textbf{Available scaffolding} & \shortstack{\textbf{Topology/build}\\\textbf{success}} & \textbf{Intent-field accuracy} & \shortstack{\textbf{Mean total}\\\textbf{tokens}} & \shortstack{\textbf{Median AI}\\\textbf{time}} \\
\hline
ScenarioForge+Qwen3.8 & Schema, compiler, catalogs, validation, repair & 50/50 (100\%) & 100\% (160/160) & 89,680 & 94.4~s \\
ScenarioForge+GPT5.6 & Same framework-assisted path & 50/50 (100\%) & 100\% (160/160) & 41,437 & 37.4~s \\
Frontier R1 & CORE XML request & 4/50 (8.0\%) & 55.0\% (88/160) & 5,812 & 76.5~s \\
Frontier R2 & R1 + structural example & 28/50 (56\%) & 85.0\% (136/160) & 2,808 & 27.1~s \\
Frontier R3 & R2 + node/service vocabulary & 38/50 (76\%) & 86.9\% (139/160) & 2,870 & 23.4~s \\
Frontier R4 & R3 + segmentation vocabulary & 38/50 (76\%) & 91.3\% (146/160) & 3,109 & 26.6~s \\
Frontier R5 & R4 + container-image guidance & 41/50 (82\%) & 91.9\% (147/160) & 3,225 & 27.1~s \\
\hline
\end{tabular}
\parbox{0.94\textwidth}{\vspace{2pt}\scriptsize EQ2 scores topology authoring; downstream Flow/challenge sequencing is derived by ScenarioForge. Buildability denotes nonempty CORE construction. Token/time values include each retained attempt's model calls.}
\end{table*}

\subsection{Matched AI-Authoring Benchmark Results}

Table~\ref{tab:frontier-comparison} answers EQ2. Both framework conditions achieved perfect buildability scores, whereas direct frontier generation improved across rungs but remained nine builds short at R5. Failure inspection adds detail beyond the aggregates: at R1, 35 artifacts loaded with no nodes, 10 were rejected, four built successfully, and one prompt had no scored artifact; all nine failures at R5 resulted from CORE rejecting the generated artifacts. Qwen and GPT tied inside the framework, indicating that model scale was not decisive under the shared schema, catalog, compiler, validation, and repair path.

Intent accuracy followed the same pattern: both framework conditions achieved perfect intent-field scores, while direct R5 remained 13 fields short and satisfied every field in 38 of 50 requests. Table~\ref{tab:frontier-comparison} also reports the cost and latency difference between the tied framework outcomes, separating authoring quality from provider effort.

The rung trajectory in Table~\ref{tab:frontier-comparison} shows that the structural example produced the largest gain; node/service vocabulary primarily improved buildability, segmentation vocabulary improved intent without changing buildability, and image guidance yielded a smaller final gain. Increasingly explicit syntax and vocabulary therefore narrowed but did not close the gap. The framework conditions achieved perfect scores on both measures without exposing this cumulative representation scaffold to the user.

All 100 framework-generated scenarios (50 per model) ran on the ARM CORE VM with Flow challenge sequencing disabled; after 60 seconds, runtime and live-artifact checks were applied. With original resources, ScenarioForge+Qwen3.8 passed 46/50 and ScenarioForge+GPT5.6 passed 47/50. Diagnostic replays recovered remaining cases: one GPT case passed unchanged after a transient CORE failure, while two GPT and four Qwen passed after documented substitutions with prior successful execution on the same ARM endpoint. The qualified replays reached 50/50 for each framework condition; first-pass and remediation outcomes remain separately reported.

The gap between 100\% construction and lower first-pass runtime/artifact success identifies the remaining failure layer: a CORE topology can construct before an architecture-incompatible container or transient daemon event interrupts startup. Reporting original resource selections and remediated replays independently distinguishes topology quality from endpoint compatibility and transient infrastructure behavior.

\subsection{Interpretation and Measurement Scope}

As Table~\ref{tab:frontier-comparison} shows, the same frontier model achieved perfect scores on both measured metrics only inside \methodname; ScenarioForge+Qwen3.8 reached them as well. Schema, catalogs, compilation, validation, and repair therefore materially change whether generated intent becomes a usable CORE definition. EQ1 and EQ2 test complementary sides of the trust boundary---description-to-definition and fixed-definition-to-validated-exercise, respectively---and neither substitutes for the other.

We used \texttt{gpt-5.6-sol} to generate the arm-neutral request suite and froze all requests and expected fields before evaluation. Intent scoring covers six observable topology and resource field types, while provider-reported token counts and latency characterize the retained authoring calls. Together with the per-case native artifacts and CORE outcomes, these measurements expose both semantic preservation and executable-structure quality.

\section{Representative Cases and Deep Dive}

Five structurally distinct cases illustrate the dataset; full records cover further compositions.

\subsection{Five-Case Comparison}

Table~\ref{tab:representative-scenarios} compares challenge structures and participant-facing evidence across varied scales, resource mixes, access prerequisites, and dependencies.

\begin{table*}[!t]
\caption{Representative scenario configurations and coverage}
\label{tab:representative-scenarios}
\centering
\scriptsize
\setlength{\tabcolsep}{3pt}
\renewcommand{\arraystretch}{0.90}
\begin{tabular}{>{\raggedright\arraybackslash\bfseries}p{0.14\textwidth} >{\raggedright\arraybackslash}p{0.25\textwidth} >{\raggedright\arraybackslash}p{0.20\textwidth} >{\raggedright\arraybackslash}p{0.34\textwidth}}
\hline
\textbf{Dimension} & \textbf{Representative run} & \textbf{Challenge structure} & \textbf{Representative artifacts} \\
\hline
Access dependencies & dataset-segmented-enterprise-pivots\_run01 & 6 total: 2 vulnerability, 4 generated; 2 access requirements & ChatOps transcript; wallet note; parser endpoint; SSH key; HTTP credential page \\
CI supply chain & dataset-mixed-ci-supply-chain\_run01 & 3 total: 1 vulnerability, 2 generated & Traversal result; \texttt{deploy-values.yaml}; SQL lookup \\
Segmented scale & dataset-scale-segmented-combined\_run05 & 10 total: 5 vulnerability, 5 generated & CSV; packet capture; encoded blobs; MD5 hash; risk register; audit dump \\
Multi-artifact access & dataset-scale-multi-artifact\_run03 & 5 generated objectives; 1 access requirement & SSH keys; lab notebook; runbook; backup/finance records; NFS evidence \\
Messaging chain & dataset-mixed-collaboration-messaging\_run03 & 4 total: 1 vulnerability, 3 generated & SMTP bounce; IMAP message; DMARC XML; SHA-256 secret \\
\hline
\end{tabular}
\end{table*}

The access-dependency case combines two vulnerability-backed services, four generated resources, and two source-access requirements. Its six-objective chain links network reachability, acquired access, and typed artifacts rather than independent flags; Sections~VI-B and VI-C examine it further.

The other cases emphasize different structures. The CI supply-chain case passes traversal evidence and \texttt{deploy-values.yaml} into a SQL lookup. The ten-step segmented-scale case combines forensic, encoded, configuration, and database evidence. The multi-artifact case links five generated objectives through an SSH access prerequisite without a vulnerability-backed objective. The messaging case uses SMTP, IMAP, and DMARC evidence to derive a SHA-256 secret. Table~\ref{tab:representative-scenarios} thus contrasts progression, access, and evidence transformation.

We examine \emph{dataset-segmented-enterprise-pivots\_run01} in depth; dataset-wide claims draw on all 226 records.

\subsection{Detailed Case: Network Placement and Segmentation}

\begin{figure*}[!t]
    \centering
    \includegraphics[width=0.96\textwidth]{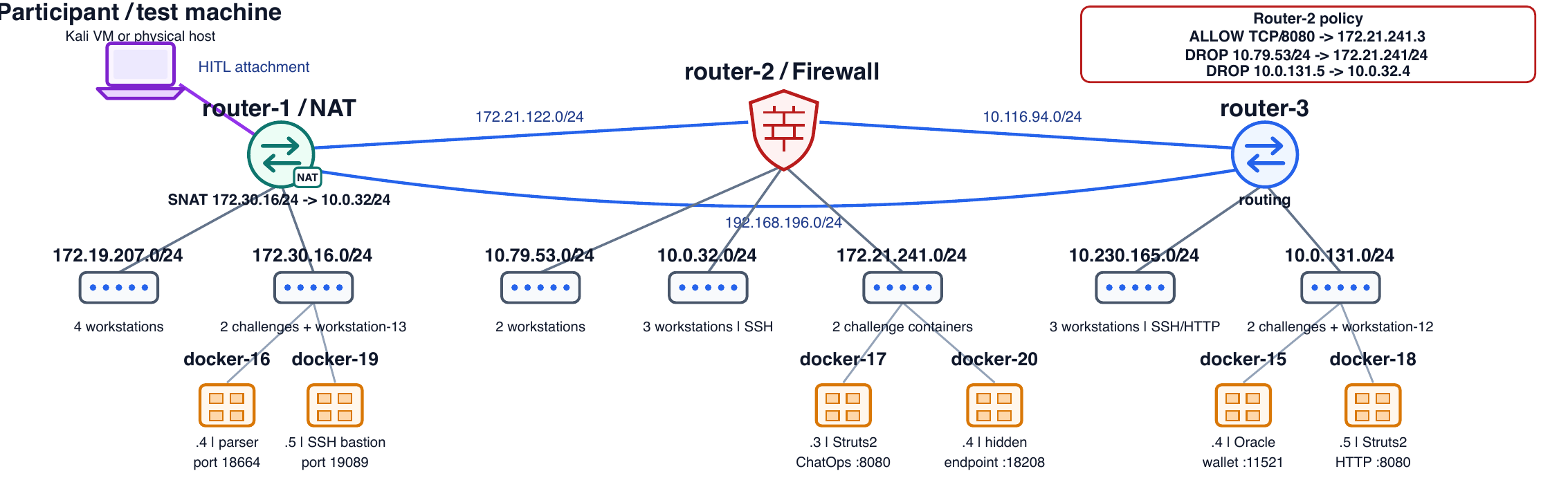}
    \caption{Representative deployment topology for \emph{dataset-segmented-enterprise-pivots\_run01}. Orange boxes identify challenge containers and service ports; purple indicates the participant's HITL attachment; red and green indicate firewall and NAT enforcement.}
    \label{fig:representative-network-topology}
\end{figure*}

Fig.~\ref{fig:representative-network-topology} shows HITL entry, NAT, firewall enforcement, and distributed challenge containers; counts distinguish hosts from challenge nodes.

\subsection{Detailed Case: Source-Access and Artifact-Dependent Challenge Graph}

\begin{figure}[!t]
    \centering
    \includegraphics[trim=41 40 41 40,clip,width=0.90\columnwidth]{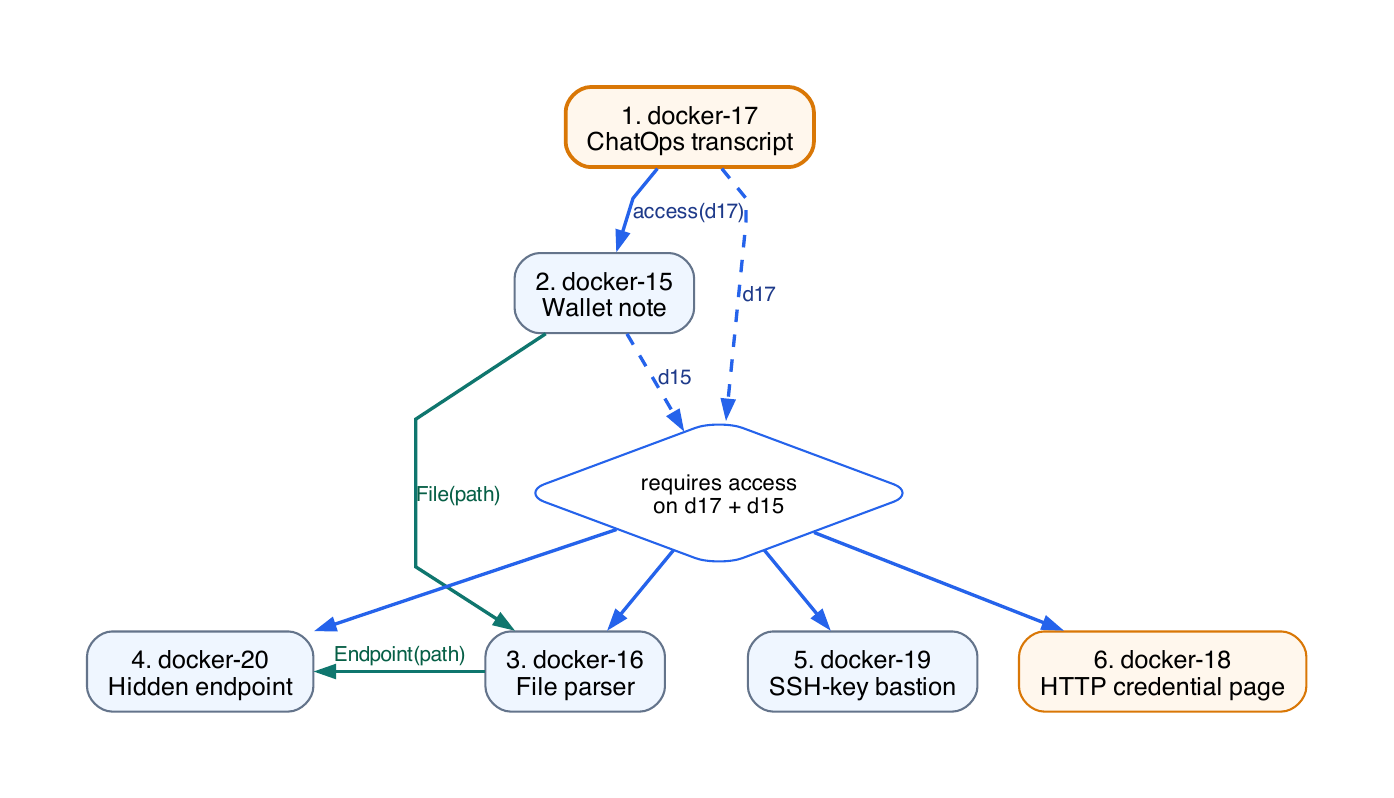}
    \caption{Source-access and artifact dependencies in \emph{dataset-segmented-enterprise-pivots\_run01}. Dashed edges show validated access prerequisites; green edges show typed artifact dependencies. Orange nodes are vulnerability-backed; blue nodes use generated resources.}
    \label{fig:representative-attack-graph}
\end{figure}

Fig.~\ref{fig:representative-attack-graph} shows the resolved dependency structure. Source access acquired on \texttt{docker-17} gates the wallet service on \texttt{docker-15}, which establishes a second access prerequisite; typed artifact edges connect the resulting wallet file to the parser and its endpoint to the follow-up. The graph therefore distinguishes access and artifact prerequisites from an enforced nested network route.

The participant enters at Router~1, reaches the policy-exposed ChatOps service on \texttt{docker-17}, and obtains the first source-access prerequisite. From that node, an Oracle wallet service on \texttt{docker-15} supplies a file and second source-access capability; the file is submitted to a parser on \texttt{docker-16}, whose returned endpoint leads to \texttt{docker-20}. Remaining objectives include an SSH-key bastion and HTTP credential page. The evaluator probes each declared source-to-target path independently rather than asserting a nested participant route. This separation of artifact dependencies, source context, and network policy prevents an obtained file or allowed route from being mistaken for completed progression.

The retained evidence operationalizes Fig.~\ref{fig:representative-attack-graph}: recorded requests and outputs show how each declared edge is exercised. Exact addresses, ports, inputs, expected outputs, and hints remain in the assignment and Solutions Script, while secret values are omitted from the paper.

\section{Release and Reproduction}

Each of the 226 Execution Dataset records includes resolved YAML, seed, authoritative XML, execution evidence, metrics, and a portable \texttt{scenarioforge-reproduction.zip} bundle. Aggregate files trace reported values to records.

The release comprises four repositories. \texttt{scenarioforge}\footnote{\url{https://anonymous.4open.science/r/scenarioforge-479A}} provides authoring, grounding, and execution. \texttt{flag-generators}\footnote{\url{https://anonymous.4open.science/r/flag-generators-E148/}} provides the challenge and Vulhub catalogs. \texttt{scenarioforge-eval}\footnote{\url{https://anonymous.4open.science/r/scenarioforge-eval-E1D7/}} supports batching, validation, and analysis. Finally, \texttt{scenarioforge-dataset}\footnote{\url{https://anonymous.4open.science/r/scenarioforge-dataset-6015/}} contains the 226 Execution Dataset records. It also contains the independent \path{frontier-comparison/} collection: 50 requests, outputs from seven conditions, scoring records, metrics, and analysis scripts. The comparison prompts are not part of the 226 execution cases.

For interactive authoring, users work through the AI Generator or WebUI, review grounded previews, and launch sequencing, execution, validation, and export. For collections, a \texttt{.spec.yaml} defines iteration count, seeds, topology and resource choices, Flow settings, segmentation, traffic, access, and validation policy. Each iteration freezes a resolved specification, enabling diverse regeneration or exact replay from authoritative XML and seed.

The command \texttt{scenarioforge-eval --sf-path SPEC\_PATH --execute --out OUTPUT\_PATH} reproduces specification collections. Replay may begin from a specification and seed or authoritative XML, bypassing model variation. The reproduction ZIP carries authoritative XML, a hashed manifest, and materialized artifact sources; WebUI import reconstructs the recorded configuration and artifact state for replay, or regenerates artifacts deterministically from saved resolved inputs. It must be protected because it may contain credentials, flags, and generated secrets. Solutions Script, HITL, and exploit-harness interfaces support participant and automated exercise interaction.

Source specifications define reusable families; resolved specifications freeze randomized choices; result directories retain XML and observed evidence. The comparison subtree preserves native responses and grades. A hashed manifest and analysis scripts support table reproduction. Connection credentials and provider secrets are redacted; replay requires separate CORE connection settings.

The release supports aggregate, per-case, and raw-evidence audits. Specifications establish EQ1's intended composition; native artifacts and request identifiers trace EQ2 outcomes. Raw XML, logs, and reports preserve negative and intermediate evidence alongside successful examples.

\section{Conclusion}

\methodname separates proposals from grounded, deployed, and validated exercises. All 226 Execution Dataset scenarios completed the full pipeline. In the independent 50-prompt Matched AI-Authoring Benchmark, both framework conditions achieved 50/50 topology/build success and 160/160 intent fields; direct frontier generation peaked at 41/50 and 147/160. The smaller open-weight model matched the frontier model inside \methodname. These complementary studies evaluate topology authoring and full-pipeline execution; the released artifacts support reproducibility and future studies comparing models, participants, and agents.

\section*{Acknowledgment}
This work was supported in part by the U.S. Army Research Laboratory under Cooperative Agreement W911NF-26-2-A210. AI tools assisted with proofreading and experimental setup; the authors remain responsible for the manuscript, methodology, and reported results.

\balance
\bibliographystyle{IEEEtran}
\bibliography{references}  
\end{document}